\documentclass[preprint,aps]{revtex4}
\usepackage{hyperref,slashed,color}
\usepackage{graphicx}
\usepackage{epstopdf}
\usepackage{multirow}
\usepackage{threeparttable}

\begin{document}
\title{The Global Electroweak Fit and its Implication to $Z'$}

\author{Ying Zhang$^1$\footnote{{\it Email address}: hepzhy@mail.xjtu.edu.cn},
Shao-Zhou Jiang$^2$\footnote{{\it  Email address}:  jsz@gxu.edu.cn},
Qing Wang$^{3,4}$\footnote{Corresponding author\\ {\it Email
address}:~wangq@mail.tsinghua.edu.cn.}}

\address{$^1$School of Science, Xi'an Jiaotong University, Xi'an, 710049, P.R.China\\
$^2$College of Physics Science and Technology, Guangxi University, Nanning, Guangxi 530004, P. R. China\\
$^3$Department of Physics, Tsinghua University, Beijing 100084, P.R.China \\
$^4$Center for High Energy Physics, Tsinghua University, Beijing 100084, P.R.China}

\date{May 26, 2012}
\begin{abstract}
 Among the $Z$-pole observables, $A_{FB}^{(0,b)}$ and $A_e$ suffer moderately-large standard deviations
 from the Standard Model predictions. Fine-tuning of the unknown Higgs mass only reduces the deviation of one of them at the expense of increasing the deviation of the other observable. If we take this fact seriously,  the result can be interpreted as  independent experimental evidence of existing new physics beyond SM, even if a 125GeV Higgs on the LHC is finally confirmed to be SM Higgs. We show in this paper that the existence of a $Z'$ boson  mixing with $Z$ and $\gamma$ and with generation-dependent anomaly-free charge assignments, helps to suppress  $A_{FB}^{(0,b)}$ and $A_e$ at $Z$-pole simultaneously and dose reduce the largest deviation from $2.7\sigma$  in SM predictions to $1.2\sigma$ in our scenario. The global electroweak fit does not prefer $Z'$ coupling to the second-generation quarks and the first-generation left-handed lepton and right-handed electron. The fitting result also supports a negative contribution to the $S$ parameter from the $Z'$ boson.


\end{abstract}
\maketitle

\section{Introduction}

Along with the successful run of LHC at 8TeV, many researchers now are anxiously waiting judgment on whether the 125GeV possible excess displayed by ATLAS and CMS experiments is the Standard Model (SM) Higgs or some other particle(s) or just a statistical fluctuation. If before the end of 2012 the excess is shown either not to be the SM Higgs or just to disappear as  more data accumulates, the SM Higgs then will be simply ignored by the LHC experiments. This will be a very cheering result, since it offers the strongest direct experiment proof that the electroweak symmetry breaking mechanism predicted by SM is not correct and new physics beyond SM is needed, despite any discovery of new physics yet. Particle physics will enter a golden age full of excitements with prospects from various investigation ventures and potential discoveries of predicted or unexpected new physics.

However, if by the end of this year, data from the ATLAS and CMS experiments support the discovery of the SM Higgs, SM then may become a perfect model not only theoretically but also experimentally. While celebrating the great success, one disadvantage that particle physicists avoid talking about, emerges more and more realistically, i.e., no new physics appears. Some have shown that, with the 125GeV Higgs, SM survives all the way to the Planck scale\cite{Shaposhnikov}. Particle physics in this scenario then has a bleak future without new phenomena emerging, apart from those predicted by SM.
The conventional basis of unclear electroweak symmetry breaking mechanism that new physics searchers relied on may no longer exist.
There are two different schemes to deal with this bad situation. The first is to insist on searching new physics in electroweak symmetry breaking sector of SM and it is already taken by many researchers. Considering that the final verification of SM Higgs will need many years of further experimental effort,  along with ever-increasing amounts of data from the experiment, one can gradually narrow various Higgs couplings to ordinary quarks, leptons and gauge bosons to their SM values to examine the possible left new physics spaces \cite{HiggsCouplings}. The second is that we propose and emphasize in this paper. Assuming the 125GeV access is exactly SM Higgs, we search new physics evidence in alternative fields instead of continuously precision check of the Higgs profile. Two apparent such facts are the existence of dark matter predicted from cosmology and massive neutrinos from their oscillation experiments. However, dark matter maybe only participates in gravitational interactions and a simple generalization of SM can introduce neutrino masses. These two evidences are not strong enough to ensure the existence of non-trivial new physics.
To raise hopes for a future for those new physics believers, we have to find some extra evidence or clues requiring new physics. We know that SM fits experiments very well up to now; based on present experimental data, we do not have strong evidence for new physics. We need to lower expectations  to investigate effects that are not so strong. Conventionally, the veracity of SM is tested by precision measurements in the electroweak sector. A well-known scheme is the global electroweak SM fit in which a series of observables at the Z-pole and off Z-pole are chosen to fit  experiment results with theoretical predictions by tuning the input parameters $\alpha$, $G_F$, $M_Z$, and other quantities which are irrelevant for this work.
The complete global fit for Z-pole observables in \cite{PDG} shows that the deviation of electroweak observables is smaller than 1$\sigma$. The exceptions
are following seven observables: $A_{FB}^{(0,b)}$ with 2.7$\sigma$, $A_e$ with 1.8$\sigma$, $\Gamma(inv)$ with 1.8$\sigma$, $\sigma_{had}$ with 1.5$\sigma$, $R_\mu$ with 1.5$\sigma$, $R_e$ with 1.4$\sigma$, and $\Gamma(had)$ with 1.3$\sigma$.
In latest improved results with Gfitter \cite{gfitter}, there are still seven observables with the deviation of more than 1$\sigma$, especially $A_{FB}^{(0,b)}$ and $R_b^0$ with the deviation of more than 2$\sigma$.
Since this is already the best fit of SM which leads unknown Higgs mass $M_H=120^{+12}_{-5}$GeV, any further variation of $M_H$ will worsen  the result. This can be further seen from $M_H$ dependence of separate fitting for above five observables. Gfitter shows that $A_{FB}^{(0,b)}$ and $A_l\mathrm{(SLD)}$ prefer smaller while $M_W$, $\sigma^0_{\mathrm{had}}$ and $R_{\mathrm{lep}}$ prefer larger $M_H$. So within SM itself, no matter what value of $M_H$, it is impossible to further improve the present  global electroweak SM fitting result.
If we insist on reducing the deviation values for these observables, the only way is to add in new physics. In this sense, the moderately large deviations in the global electroweak SM fit are the evidences that we seek re-demonstrate the existence of new physics. Improving   global electroweak SM fits can then provide a measure to judge  the new physics models.

In the literature, we do not find many works treating the global electroweak SM fits in this way.
One possible reason is that the deviation of the fit is not so large, and researchers have turned their attention to other observables, such as the oblique parameters $S$, $T$ and $U$ \cite{gfitter}, especially $S$ which often gives very strong constraints on  new physics models. In fact, many years ago, when $R_b$ received a very large deviation \cite{RbHistory}, much  work appeared trying to explain the result in terms of various new physics models.  Along with improvements in experimental results, the deviation in $R_b$ was reduced over time; researchers were then  no longer interested in those kinds of discussions. Now, assuming that the SM Higgs is confirmed,  what we want to do is to re-initiate the investigation in this direction. With this  application of global electroweak fitting to new physics models in mind, we examine below the simplest Abelian extension of SM that can be used as an example of the new physics beyond SM. In this scenario, the conventional SM gauge group $SU(3)_c\otimes SU(2)_L\otimes U(1)_Y$ is extended to $SU(3)_c\otimes SU(2)_L\otimes U(1)_Y\otimes U(1)'$. The neutral heavy gauge boson associated with the extra broken $U(1)'$ group is usually called the $Z'$ boson, assumed here to be the lightest new physics particle beyond SM. Many new physics models have one such $Z'$ boson (for details see Ref.\cite{LangackerArxiv2008,LangackerPRL2008}) as a  remnant of new physics interactions. For this paper, we take $Z'$ as a simplest case of new physics to study its effects on the global electroweak fit. We find  three references in the literature that investigated such effects of a $Z'$ particle \cite{fit1,fit2,Z'Zpole}. In Ref.\cite{fit1}, the author performed a complete electroweak fits to Z' extensions of SM and pointed out that presence of $Z'$ may solve $A_{FB}^b$ anomaly problem. In Ref.\cite{fit2}, it was shown that a $Z'$ based on gauged $L_{\mu}-L_{\tau}$ lepton number can improve the fit for the anomalous magnetic moment of the muon. In Ref.\cite{Z'Zpole}, the author claimed that a $Z'$,  with suppressed couplings to the electron compared to the $Z$ boson and couplings to the $b$ quark, provides an excellent fit to $A_{FB}^{(0,b)}$ and $A_e$ and significant improvements in $\sigma^0_{\mathrm{had}}$ and $R_b$ at $M_Z+$2GeV scale. To make a relatively large contribution, the  $Z'$ mass was fitted close to $Z$ boson mass, otherwise the contribution from $Z'$ exchange vanishes unless $Z'$ is near the $Z$-pole. In Ref.\cite{fit2}, only the simplest $Z-Z'$ mixing was considered, while in the analysis of Refs.\cite{fit1,Z'Zpole}, effect from $Z-\gamma-Z'$ mixing was ignored.  This is OK for the fittings for those  observables out of Z-pole, where mixing effect is small. But for the Z-pole observables we focused in this paper, fitting result is sensitive to the $Z-\gamma-Z'$ mixings. To avoid the unnatural $Z'$ mass near $M_Z$ and include in important $Z-\gamma-Z'$ mixing effect, we consider the most general model-independent $Z'$ boson described by an electroweak chiral Lagrangian given in our previous paper \cite{OurJHEP2008}. In fact, from an effective field theory point of view, all coefficients appearing in the electroweak chiral Lagrangian \cite{OurJHEP2008} are free parameters which need to be fixed either from experiments or from underlying theoretical models. Our global fit can be seen as a kind of procedure to extract out experiment constraints on these coefficients. In  contrast to Ref.\cite{Z'Zpole}, in this fit the $Z'$ mass needs not to  close to Z-pole due to some extra free parameters not fixed by the global fit. Direct exchange at the $Z$-pole discussed in Ref.\cite{Z'Zpole} then becomes enough small to be neglected in our work.
Other $Z'$ bosons effects include modifications of $Z$ couplings through $Z-\gamma-Z'$ mixings and $Z'$ couplings to SM fermions. They are the main sources to cause the change. In next section, we will discuss how to manipulate them and obtain our improved global electroweak fitting result.

\section{Global Electroweak at Z-pole with inclusion of $Z'$}

This is the main part of the paper. We will first discuss $Z-\gamma-Z'$ mixings and $Z'$ couplings to SM fermions separately, then describe our global fitting procedure. Finally give our fitting result and make some discussions.

 For $Z'$ mixings with electroweak neutral bosons $Z$ and $\gamma$, usually a lighter $Z'$ is possible at larger mixing angles; smaller mixing angles arise only for heavier $Z'$ (see Ref.\cite{HewettPR1989}). More detailed results depend on the form of the mixings  set by the models. Considering that the experimental constraint for  the $Z'$ mass has already reached the TeV energy region, we assume here a small mixing (order of $10^{-3}$) $Z'$. Based on a model independent chiral Lagrangian, we have already classified and parameterized the most general $Z-\gamma-Z'$ mixings in Ref.\cite{OurJHEP2009,OurCPC2012}. To include in $Z'$ mixings effects, we define the rotation matrix between  the gauge eigenstates
 of the neutral gauge bosons of $SU(2)_L\otimes U(1)_Y\otimes U(1)'$ and the corresponding mass eigenstates as follows:
\begin{eqnarray}
	\left(\begin{array}{c}W^3_\mu\\ B_\mu\\ X_\mu\end{array}\right)
		=U\left(\begin{array}{c}Z_\mu\\ A_\mu\\ Z'_\mu\end{array}\right)\hspace{2cm}
	U=\left(\begin{array}{ccc}
		c_W+\Delta_{11}& s_W+\Delta_{12} &  \Delta_{13}
		\\
		-s_W+\Delta_{21} & c_W+\Delta_{22} & \Delta_{23}
		\\
		\Delta_{31} & \Delta_{32} &1+ \Delta_{33}
		\end{array}\right)\label{Umatrix}
\end{eqnarray}
with Weinberg angle $\theta_W$ and $s_W=\sin\theta_W,~c_W=\cos\theta_W$.
After rotation by $U$, the neutral gauge fields will be diagonalized into their mass eigenstates.

For the $Z'$ coupling to SM fermions, we need to determine the $U(1)'$ charge assignments of the SM fermions.  These are related to following neutral current interactions
\begin{eqnarray}
-\mathcal{L}_{NC}&=&gW^3_\mu J^{3,\mu}+g'B_\mu J^\mu_Y+{g''}X_\mu J^\mu_X\;,
\nonumber
\end{eqnarray}
 where $g$,$g'$ and $g''$ are the respective $SU(2)_L$, $U(1)_Y$ and $U(1)'$ gauge coupling constants and
\begin{eqnarray}
J^{\mu}_3&=&\sum_i\bar{f}_i\gamma^\mu t_{3iL}P_Lf_i
\nonumber\\
J^\mu_Y&=&\sum_i\bar{f}_i\gamma^\mu[y_{iL}P_L+y_{iR}P_R]f_i
\nonumber\\
J^\mu_X&=&\sum_i\bar{f}_i\gamma^\mu[y'_{iL}P_L+y'_{iR}P_R]f_i\label{JXdef}
\end{eqnarray}
are neutral currents corresponding respectively to the third component gauge boson $W^3_\mu$ of weak isospin, hypercharge gauge boson $B_\mu$ and an extra $U(1)'$ gauge boson $X_\mu$. In addition, $y'_{iL,R}$ are the left/right-handed $U(1)'$ charges for the SM fermion field $f_i$ with flavor index $i$;  the electroweak gauge symmetry $SU(2)_L$ constrains these to eighteen charges in the most general case: three left-handed quark charges for each of the three generations $y_{q_1}'$, $y_{q_2}'$, $y_{q_3}'$; six right-handed quark charges for each quark $y_u'$, $y_d'$, $y_c'$, $y_s'$, $y_t'$, $y_b'$; three left-handed lepton charges for each generation $y_{l_1}'$, $y_{l_2}'$, $y_{l_3}'$; and six right-handed lepton charges for each leptons $y_e'$, $y_{\nu_e}'$, $y_{\mu}'$, $y_{\nu_\mu}'$, $y_{\tau}'$, $y_{\nu_\tau}'$. In the mass eigenstates basis, the neutral currents become
\begin{eqnarray}
-\mathcal{L}_{NC}&=&e^*J_{em}^\mu A_\mu+g_ZJ_Z^\mu Z_\mu+{g''}J_{Z'}^\mu Z'_\mu\nonumber
\end{eqnarray}
with $e^*$ being the renormalized electric charge incorporating  $Z'$ contributions and $g_Z=\sqrt{g^2+{g'}^2}$. For the rotation matrix (\ref{Umatrix}), the neutral currents in the mass eigenstates basis are related to those in the interaction eigenstates basis as:
\begin{eqnarray}
e^*J_{em}^\mu&=&g(s_W+\Delta_{12})J^{3,\mu}+g'(c_W+\Delta_{22})J^{\mu}_Y+{g''}\Delta_{32}J^\mu_X
\label{Jem}\\
g_ZJ^\mu_Z&=&g(c_W+\Delta_{11})J^{3,\mu}+g'(-s_W+\Delta_{21})J^{\mu}_Y+{g''}\Delta_{31}J^\mu_X
\label{Jz}\\
{g''}J_{Z'}^\mu&=&g\Delta_{13}J^{3,\mu}+g'\Delta_{23}J^{\mu}_Y+{g''}(1+\Delta_{33})J^\mu_X.
\label{Jzprime}
\end{eqnarray}
To keep the gauge symmetry, the assignments of the charges must satisfy the anomaly cancelation conditions.
A generation independent solution has been discussed in our previous paper \cite{OurJHEP2009}, in which only two charges are free due to cancelations of $[SU(3)_C]^2U(1)'$, $[SU(2)_L]^2U(1)'$, $U(1)_Y[U(1)']^2$, $[U(1)_Y]^2U(1)'$, $[U(1)']^3$ and the mixing gravitational-gauge anomalies.
To obtain the most general fitting result, we focused on making  general generation dependent charge assignments. Considering that some charges are very small in the fitting result, we further discuss a case in which these small charges vanish. In all,  two cases arise:
 \begin{enumerate}
\item[Case 1:] The anomaly  cancels only when all three generations are included. We find that there are twelve  free charges.
\item[Case 2:] Within Case 1, we further stipulate that the charges for the second-generation quarks, the first-generation  left-handed lepton, and right-handed electron vanish, i.e.
$y'_{q_2}=y'_{c}=y'_{s}=y'_{l_1}=y'_{e}=0$. Six free charges exist.
\end{enumerate}

To simplify the fitting procedure and give prominence to our physical result, we performed a global electroweak fit by fixing the SM contribution with its  predictions given by PDG2010 \cite{PDG}, and only investigated the possible $Z'$ tree order contributions. A more subtle alternative fit without fixing SM contributions, which will need more complex SM computations, will be discussed  elsewhere. To realize this simplified global electroweak fit, we took three precisely measured observables, $\alpha$, $G_F$ and $M_Z$,  as input parameters with which to express the three SM three parameters: $g$, $g'$ and electroweak vacuum expectation value $v$ in the SM Lagrangian. Other SM parameters, such as fermion masses $m_f$, Higgs mass $m_H$, and strong coupling constant $\alpha_s$, are irrelevant in the fit to fix SM contributions.

\begin{figure}[htbp]
\centering
\includegraphics[scale=0.4]{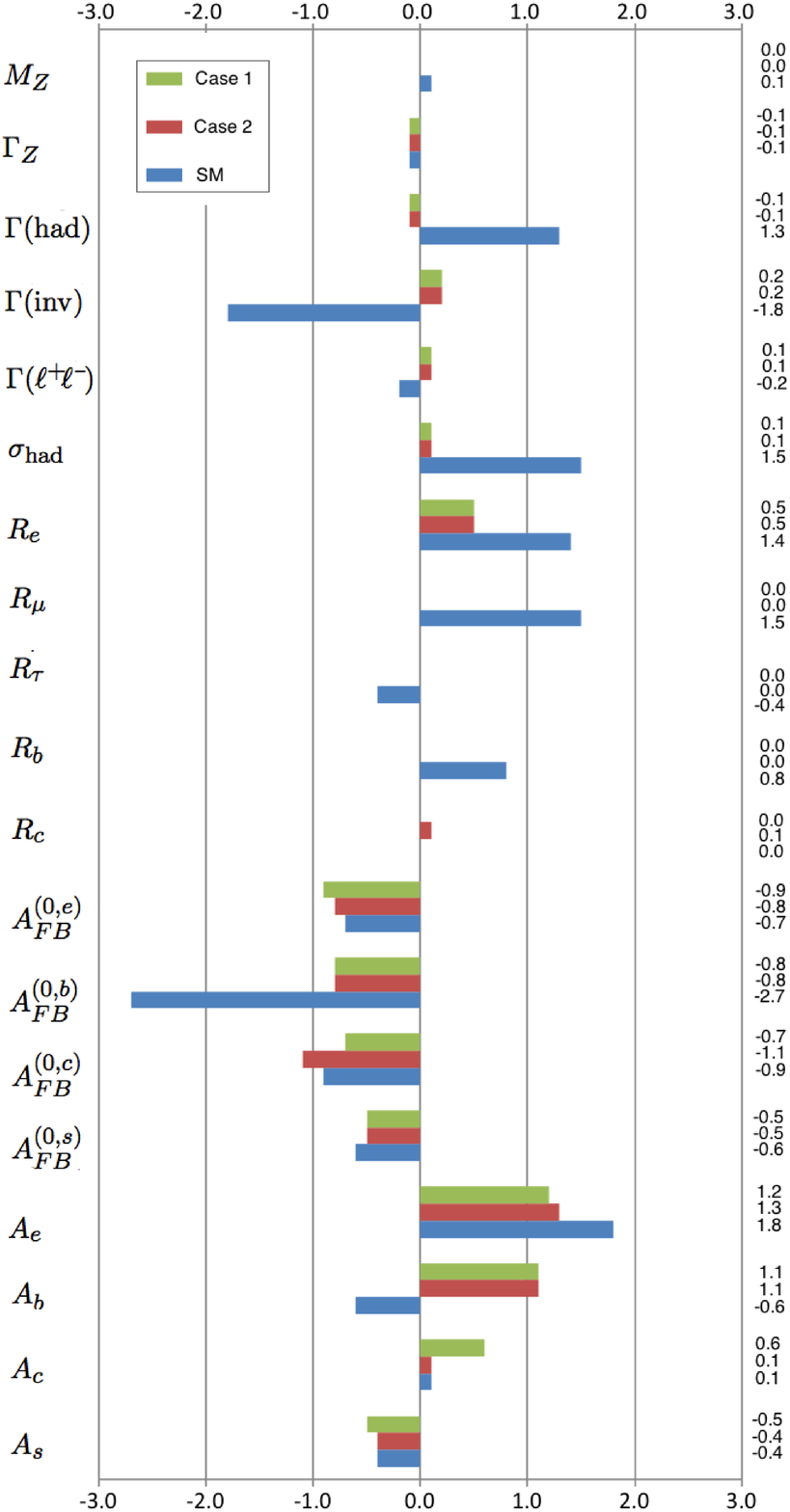}
\caption{$Z'$ pull}
\label{Fig.pull}
\end{figure}

At tree level, we fix $G_F$. The fine structure constant $\alpha=\frac{{e}^2}{4\pi}$ and Z boson mass $M_Z$ are corrected by modifications including the respective  contributions to $\alpha^*=\alpha(1+{\Delta_{12}/s_W})^2$ and $M_Z^{*2}$
\begin{eqnarray}
M_Z^{*2}
	=M_Z^2\left\{
	\Big(1+(c_W\Delta_{11}-s_W\Delta_{21})\Big)^2+\Big(\sqrt{1-2\beta_3}\frac{g''c_Ws_W}{e}\Delta_{31}\Big)^2\right\}\;~~~~~\label{ZMass}
\end{eqnarray}
with vacuum expectation value $f$  and $Z$ mass $M_Z=\frac{f}{2}\frac{e}{c_Ws_W}$ in SM. An electroweak observable $\mathcal{O}_{\mathrm{th}}$, depending on $G_F$, $\alpha^*$ and $M_Z^*$, can be divided into two parts: one is $\mathcal{O}_{\mathrm{SM}}(G_F,\alpha,M_Z)$ coming from SM fitting values in \cite{PDG} and depending on $G_F$ and SM $\alpha$ and $M_Z$,  and the other is $\Delta\mathcal{O}_{Z'}(\Delta_{ij},y'_i)$ coming from a $Z'$ correction depending on mixing parameters $\Delta_{ij}$ introduced in (\ref{Umatrix}),  SM fermion $U(1)'$ charges $y'_i$ and constrained by the anomaly cancelation condition, i.e.,
\begin{eqnarray*}
\mathcal{O}_{\mathrm{th}}(G_F,\alpha^*,M_Z^*)=\mathcal{O}_{\mathrm{SM}}(G_F,\alpha,M_Z)+\Delta\mathcal{O}_{Z'}(\Delta_{ij},y'_i).
\end{eqnarray*}
Note, some of $\Delta_{ij}$ such as $\Delta_{13}$, $\Delta_{23}$, $\Delta_{32}$ and $\Delta_{33}$ do not enter into the formulae to those Z-pole electroweak observables. Therefore these $\Delta_{ij}$s are irrelevant to our global fit and we will not discuss their values in this paper.
The difference between the present experimental data and SM fitting results will provide a narrow space for the $Z'$ correction $\Delta\mathcal{O}_Z$. We solve for the minimum of $\chi^2$ function
\begin{eqnarray*}
	\chi^2=\sum_{\mathrm{observables}}\left(\frac{\mathcal{O}_{\mathrm{exp}}-\mathcal{O}_{\mathrm{th}}}{\delta\mathcal{O}}\right)^2
=\sum_{\mathrm{observables}}\left(\frac{\mathcal{O}_{\mathrm{exp}}-(\mathcal{O}_{\mathrm{SM}}+\Delta\mathcal{O}_{Z'})}{\delta\mathcal{O}}\right)^2
\end{eqnarray*}
by tuning $\Delta_{ij}$ and $y'_i$ to obtain our global electroweak fitting results, the smaller the $\chi^2$, the better the results. Here, $\mathcal{O}^i_{\mathrm{exp}}$ and $\delta\mathcal{O}^i$ are experimental values and errors, respectively.
We consider all $Z$-pole observables listed in \cite{gfitter} and \cite{PDG}; their definitions and expressions at $Z$-pole appear in the literature \cite{LEP2006}. The fitting result on the mixing parameters and SM fermion $U(1)'$ charges is given in Table.\ref{TAB:parameters}.
\begin{table}[htdp]
\caption{Fitting result on the mixing parameters $\Delta_{ij}$ and SM fermion $U(1)'$ charges \hspace*{2cm}in $Z'$ effective theory.}
\label{TAB:parameters}
\centering
\begin{threeparttable}
\begin{tabular}{c|c|c}
\hline\hline
 & case 1 & case 2
\\
\hline
$\chi^2$ \tnote{1}~~ & $5.7$ & $6.1$
\\
\hline
$\Delta_{ij}[10^{-3}]$ & $\begin{array}{l}\Delta_{11}=8.1\\ \Delta_{12}=-4.3\\ \Delta_{21}=-3.9\\ \sqrt{1-2\beta_3g''\Delta_{31}}=3.4\tnote{2}\\ \Delta_{22}=-7.8\tnote{3}\end{array}$
	& $\begin{array}{l}\Delta_{11}=-0.042\\ \Delta_{12}=-0.15\\ \Delta_{21}=0.40\\ \sqrt{1-2\beta_3}g''\Delta_{31}=4.5\\ \Delta_{22}=-0.27\end{array}$
\\
\hline
$y'_i$ & $\begin{array}{ll}y'_{q_1}=0.90, & y'_{l_1}=0.53,
			\\
			y'_{u}=6.1, &y'_e=-0.44,
			\\
			y'_{d}=-6.1, &y'_{\nu_e}=-4.7,
			\\
			y'_{q_2}=-0.95, &y'_{l_2}=-2.6,
			\\
			y'_{c}=-0.43,& y'_{\mu}=-4.4,
			\\
			y'_{s}=-0.43,& y'_{\nu_\mu}=2.5,
			\\
			y'_{q_3}=1.4, &y'_{l_3}=-1.9,
			\\
			y'_{t}=-0.22, & y'_{\tau}=-3.3,
			\\
			y'_{b}=3.8, & y'_{\nu_\tau}=2.3;
			\end{array}$
			& $\begin{array}{ll}y'_{q_1}=0.18,& y'_{l_1}=0,
			\\
			y'_u=-1.7,&y'_e=0,
			\\
			y'_d=-2.9,& y'_{\nu_e}=0.93,
			\\
			y'_{q_2}=0, & y'_{l_2}=1.6,
			\\
			y'_c=0,& y'_\mu=1.7,
			\\
			y'_s=0,& y'_{\nu_\mu}=2.0,
			\\
			y'_{q_3}=0.42, & y'_{l_3}=-3.4,
			\\
			y'_t=3.0, & y'_\tau=-4.2,
			\\
			y'_b=2.9, &y'_{\nu_\tau}=-4.1.
			\\
			\end{array}$
\\
\hline\hline
\end{tabular}
\begin{tablenotes}
\footnotesize
	\item[1] The value of SM $\chi^2$ is $24.5$.
	\item[2] $\Delta_{31}$ is not decided by fit independently. $\beta_3$ is the coefficient of  $\frac{f^2}{4}(tr[\hat{V}_\mu])^2$ in chiral effective theory given in Ref.\cite{OurCPC2012}.
	\item[3] It comes from the relation $c_W\Delta_{12}=s_W\Delta_{22}$ to keep photon massless.
\end{tablenotes}
\end{threeparttable}
\end{table}%
 The global fit result is given in Fig.\ref{Fig.pull}.
We see that the best fit is Case 1, for which the largest two deviations in SM fit, $A_{FB}^{(0,b)}$ and $A_e$, are obviously reduced. The largest two deviations in Case 1 is 1.2$\sigma$ for $A_e$ and 1.1$\sigma$ for $A_b$, and deviations for other observables are below $1\sigma$.
This implies that the global fit of the electroweak observables does support Case 1. 
The $Z'$ mass cannot be fixed in our fit due to following relation
\begin{eqnarray}
M_{Z'}^2=f^2[g''^2(1-2\beta_3)(1+\Delta_{33})^2+\frac{1}{4}(1-2\beta_1)(g\Delta_{13}-g'\Delta_{23})^2]
\end{eqnarray}
with $f=246$GeV and the coefficient $\beta_1$ of the term $\frac{f^2}{4}(tr[T\hat{V}_\mu])^2$ in chiral effective theory \cite{OurCPC2012}. 
Due to the undetermined parameters $g''$, $\Delta_{23}$, $\Delta_{33}$, and $\beta_3$, $M_{Z'}$ keeps free in our result. Note heavy $Z'$ with $M_{Z'}/f\gg 1$ leads large negative $\beta_3$ for small couplings $g''$ and small mixing parameters $\Delta_{23}$ and $\Delta_{33}$.

Considering that Case 1 predicts that some charges take very small values, we annul these charges by further assuming in Case 2 null charges for the second-generation quarks, the first-generation  left-handed lepton, and right-handed electron.  For Case 2, the largest deviation is still $1.2\sigma$ for $A_b$ with $\chi^2$ value of $6.1$ and we find relatively large couplings to the third generation which is matched with result of Ref.\cite{fit1}. Of course, one can continuing annulling more charges, but we find then either the absolute value of the largest deviation is larger than $1.4\sigma$--half of the largest deviation for SM, or we may have  $Z'$ mixings of order $O(10^{-2})$. We drop these cases anymore, because in further loosening the deviation value, the largest absolute value becomes $2.7\sigma$ signifying that all charges vanish and we have just reverted to SM. Large $Z'$ mixings may yield light $Z'$ \cite{HewettPR1989} or contradict with electroweak precision data. It should be noted that the mixing parameters $\Delta_{ij}$ used in this paper are not and usually are larger than the conventional mixing  angle used in literature. The conventional $Z-Z'$ mixing angle \cite{PDG} is defined $Z-Z'$ mixing angle as
\begin{eqnarray}
\tan^2\theta'=\frac{\delta M^2_Z}{M_{Z'}^2-M_{Z_0}^2}
\simeq\frac{[(c_W\Delta_{11}-s_W\Delta_{21})^2+(1-2\beta_3)(\frac{c_Ws_W}{e}g''\Delta_{31})^2]M_Z^2}{M_{Z'}^2-M_Z^2}.\label{theta'}
\end{eqnarray}
Here, $M_{Z_0}=\frac{ef}{2c_Ws_W}$ is SM $Z$ mass and $\delta M_Z^2$ stands for mixing correction to $Z$ mass. In our fit, $\delta M_Z^2$ is the order of less than $10^{-2}\mathrm{GeV}^2$ (corresponding to $Z$ mass pull less then $0.1$). Then the order of $\theta'$ is $10^{-5}$, $10^{-4}$ and $10^{-3}$ for $M_{Z'}\sim$ 10TeV, 1TeV and $100$GeV respectively. And in Case 1 and 2 the numerator of (\ref{theta'}) $[(c_W\Delta_{11}-s_W\Delta_{21})^2+(1-2\beta_3)(\frac{c_Ws_W}{e}g''\Delta_{31})^2]$ always is  $10^{-6}$, which just matches our fitting results given in Table.\ref{TAB:parameters}.

 Except the magnitudes for the mixing parameters $\Delta_{ij}$, in Case 2, there appear some null charges which implies that $Z'$ does not couple to the corresponding fermions.  The implications of these decouplings on the new physics model building need further investigations.

With the above global fitting results, we can further check effects on the oblique parameter $S$. The experiments prefer small  negative values, whereas many unsuccessful new physics models contribute it positive corrections. Thus we have  a test for our global fit.  The $Z'$ correction  contribution to $S$ is
\begin{eqnarray}
\Delta S&=&\frac{4s_Wc_W}{\alpha}\bigg[s_W\Delta_{11}-2s_Wc_W(s_W\Delta_{12}+c_W\Delta_{22})-c_W\Delta_{21}\bigg]\label{S}
\end{eqnarray}
for which we  find a contribution of $-0.05$ for Case 1 and  $-0.03$ for Case 2. i.e., our fitting result does push $S$ parameter to its experiment preferred negative region.

\section{Summary and discussions}

	To summarize, we pointed out and emphasized  that even if LHC experiments would confirm SM Higgs, the global electroweak fit always implies the existence of new physics. Mixing of a hypothetical $Z'$  with $Z$ and $\gamma$ at order of $10^{-3}$ and couples to SM fermions with suitable anomaly free charge assignments will really improve global fitting results. It can change $\chi^2$ substantially from $24.5$ of SM to $5.7$, reduce the largest deviation from $2.7\sigma$ in SM to $1.2\sigma$ in our scenario,  suppress all observables below $1\sigma$ except $A_e$ and $A_b$, and make $Z'$ extra contributions to the oblique parameter $S$ to $-0.05$. Global electroweak fits do not prefer $Z'$ couplings to the second-generation quarks, the first-generation  left-handed lepton, and right-handed electron. 

Our global fit is done for Z-pole observables which is sensitive to $Z-\gamma-Z'$ mixing and leads typical correction at order of $10^{-3}$. This is only the part of the most general global fitting program which includes the fits for both Z-pole and off Z-pole observables, but any of new physics models must at least match Z-pole observables. Usually the new particles beyond SM are heavier than $M_Z$, which implies that their exchange effects at Z-pole are small. To more detail analyze, the correction is at most at order of $M_Z\Gamma_Z/\Lambda_{NP}^2\sim 10^{-4}$, with typical new physics particle mass $\Lambda_{NP}$ at order of 1TeV. This is one order of magnitude smaller than the correction from our $Z-\gamma-Z'$ mixing. Therefore if there are not enough big mixings among new physics particles and SM electroweak gauge particles, we cannot expect generally sufficient large improvements on global electroweak fit at Z-pole from
exchange of new physics particles. Among various new physics particles, the most important particle which mixes with the SM electroweak gauge particle is $Z'$. The other possible charged vector particles $W^{\pm\prime}$ are less important due to their relative simple mixings with $W^{\pm}$, Ref.\cite{fit1} gives the simliar result. Further the loop correction from these new physics particles usually suppressed by loop factor $1/16\pi^2$ which may be at the same order of our $Z-\gamma-Z'$ mixing depending on the detail value of coupling of the loop.
For the global fit coming from the loop corrections of new physics particles and from those observables off the Z-pole scale, we will leave the our corresponding discussions in future investigations.

\section*{Acknowledgment}
This work was supported by National Science Foundation of China (NSFC) under Grant No.11147192, 11075085 and 11005084, Specialized Research Fund for the Doctoral Program of High Education of
China No.20110002110010, Fundamental Research Funds for the Central Universities, and Scientific Research Foundation of GuangXi University Grant No. XBZ100686.

\end{document}